%% file: main.tex
\documentclass[final,3p,times,twocolumn]{elsarticle}
\usepackage[english]{babel}
\usepackage{amsmath, graphicx, xcolor, subcaption, hyperref, url}

\usepackage[utf8]{inputenc}

\newcommand{\cit}[1]{\textsuperscript{\cite{#1}}}

\journal{}

\begin{document}

\begin{frontmatter}

\title{AWESAM: A Python Module for Automated Volcanic Event Detection Applied to Stromboli}

\include{authors}

\begin{abstract}
Many active volcanoes in the world exhibit Strombolian activity, which is typically characterized by relatively frequent mild events and also by rare and much more destructive major explosions and paroxysms. Detailed analyses of past major and minor events can help to understand the eruptive behavior of the volcano and the underlying physical and chemical processes. Catalogs of volcanic eruptions may be established using continuous seismic recordings at stations in the proximity of volcanoes. However, in many cases, the analysis of the recordings relies heavily on the manual picking of events by human experts. Recently developed Machine Learning-based approaches require large training data sets which may not be available a priori. Here, we propose an alternative automated approach: the \textbf{A}daptive-\textbf{W}indow Volcanic \textbf{E}vent \textbf{S}election \textbf{A}nalysis \textbf{M}odule (AWESAM).  This process of creating event catalogs consists of three main steps: (i) identification of potential volcanic events based on squared ground-velocity amplitudes, an adaptive MaxFilter, and a prominence threshold. (ii) catalog consolidation by comparing and verification of the initial detections based on recordings from two different seismic stations. (iii) identification and exclusion of signals from regional tectonic earthquakes. The software package is applied to publicly accessible continuous seismic recordings from two almost equidistant stations at Stromboli volcano in Italy. We tested AWESAM by comparison with a hand-picked catalog and found that around 95~\% of the eruptions with a signal-to-noise ratio above three are detected. In a first application, we derive a new amplitude-frequency relationship from over 290.000 volcanic events at Stromboli during 2019-2020. The module allows for a straightforward generalization and application to other volcanoes worldwide.
\end{abstract}


\end{frontmatter}

\section{Introduction}

\begin{figure}[tb]
    \includegraphics[width=\columnwidth]{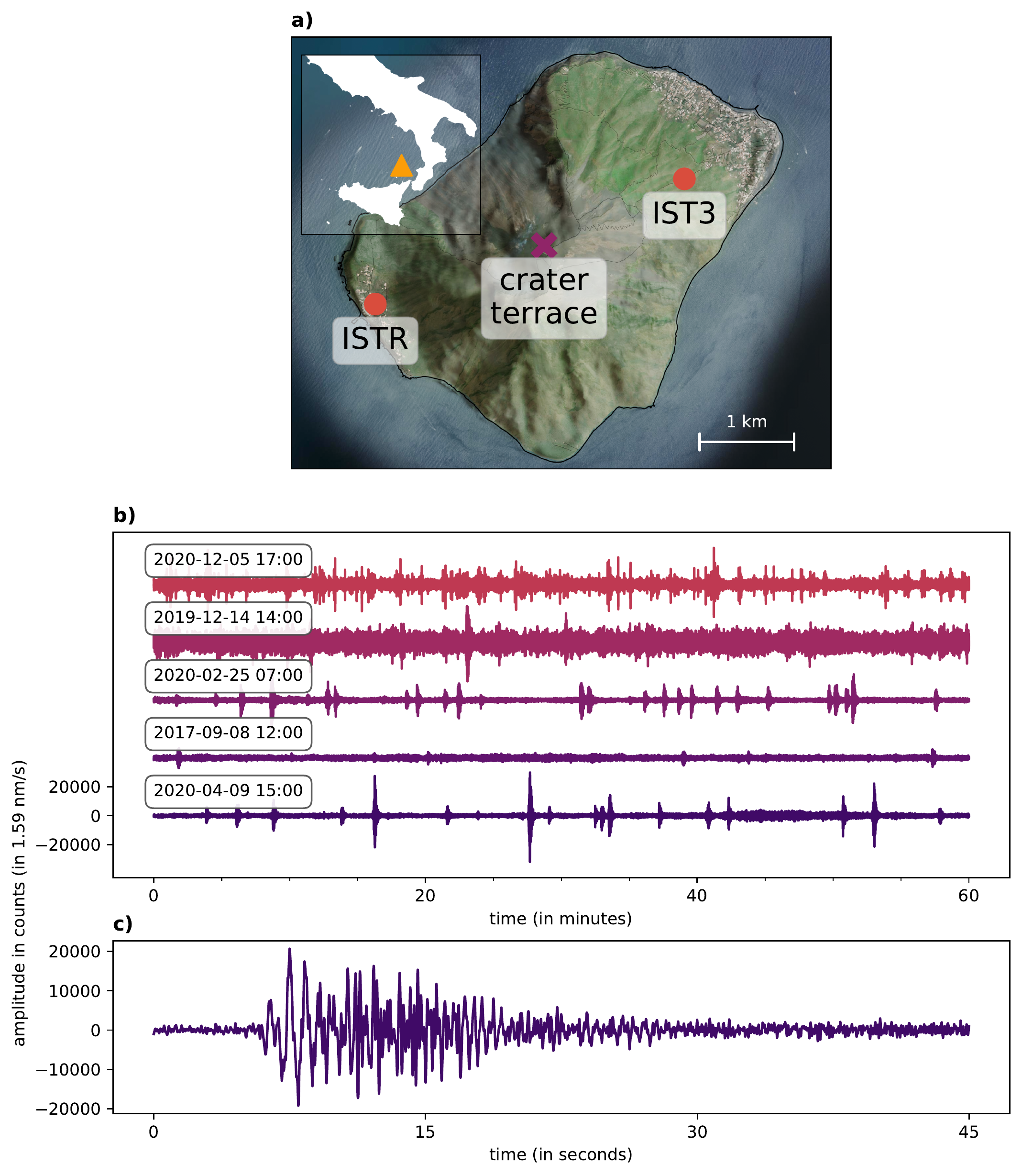}
    \caption{(a) IST3 and ISTR are the two almost equidistant seismic stations on Stromboli used for the detection of volcanic events (topographic data from openstreetmap.org\cit{osm}). 
    (b) Example seismic data with different noise levels and inter-event times recorded at station IST3, channel HHN.
    (c) Typical waveform of a single explosive VPL (very long period) event recorded at IST3, HHN. }
    \label{fig:ExData}
\end{figure}

The creation of volcanic event catalogs plays an important role in the monitoring of volcanoes. For example, studies to derive amplitude-frequency relations and inter-event time characteristics from catalogs provide important insights in understanding the eruptive behavior of volcanoes. It not only enables the evaluation of volcanic hazard, but also the in-depth investigation of possible precursors based on the historic data. Furthermore, these studies are also essential to develop and test early warning systems. For volcanoes with frequent explosive events, e.g. Strombolian volcanoes, compiling these catalogs is time-consuming and challenging as it often involves manual work from experts. In this study, we focus on Stromboli volcano in the Mediterranean Sea (Fig. \ref{fig:ExData}a). However, there are numerous volcanoes worldwide that exhibit similar Strombolian activity such as Sakurajima in Japan, Semeru in Indonesia, as well as Yasur in Vanuatu, to name just a few \cit{gutrichter, lowfreqnoise}. The methods and code presented here are, therefore, applicable to a number of additional cases to be investigated in future studies.

Volcanic eruptions (mild ones, but also paroxysms) cause ground movements similar to earthquakes that can be recorded with seismometers for monitoring (Fig.~\ref{fig:ExData}b). There are well-developed dense seismic networks around many volcanoes which are used by researchers to generate event catalogs of volcanic explosions, usually referred to as VLP (very long period) events\cit{firenze}. However, the handling of such big databases to extract the information about the events from the seismic data is a challenge, because it heavily relies on manual event picking.\cit{vulcanears}

Several automatic methods to extract volcanic explosion events from seismic data have been described previously. The short-term-average/long-term-average (STA/LTA) algorithm\cit{stalta} is a classical trigger algorithm to detect sudden changes in the data. The downside to this method is that it will not work reliably for events with a slowly increasing amplitude and its performance depends heavily on the choice of the following three parameters: short-term and long-term window length, as well as the trigger threshold.

Machine Learning based approaches, for instance Support Vector Machines (Masotti et al., 2006\cit{svm}), Hidden Markov Models (Benitez et al., 2009\cit{hmm}), and Deep Learning Models (Cortés et al., 2021\cit{vulcanears}) may also be applied. The drawback of these methods is that they require large labeled training, testing and validation datasets. These datasets must be created by other means, which may be time-consuming and labor-intensive. Another challenge for methods based on Machine Learning is the reliable detection of large events, since they are also extremely rare in the training data sets.

In order to overcome these drawbacks, we introduce the \textbf{A}daptive-\textbf{W}indow Volcanic \textbf{E}vent \textbf{S}election \textbf{A}nalysis \textbf{M}odule (AWESAM) to generate catalogs with volcanic explosion events. Fig.~\ref{fig:pipeline} gives an overview of the whole process. The strength of this algorithm is its ability to detect rare very large events, as well as frequent small events. Furthermore, the algorithm also includes a method for validating the catalog with the help of a second station. In this step, which we call catalog consolidation, local disturbances that only occur at one of the stations, and may lead to false detections, are excluded by comparison of the recordings at the two stations. This step is also used to identify and exclude tectonic earthquakes from the catalog. Altogether, AWESAM is a powerful module to create more complete volcanic event catalogs with the time and amplitude information extracted from unlabeled raw seismic waveforms.

\section{Compilation of volcanic event catalogs}
\begin{figure}[tb]
    \centering
    \includegraphics[width=0.8\columnwidth]{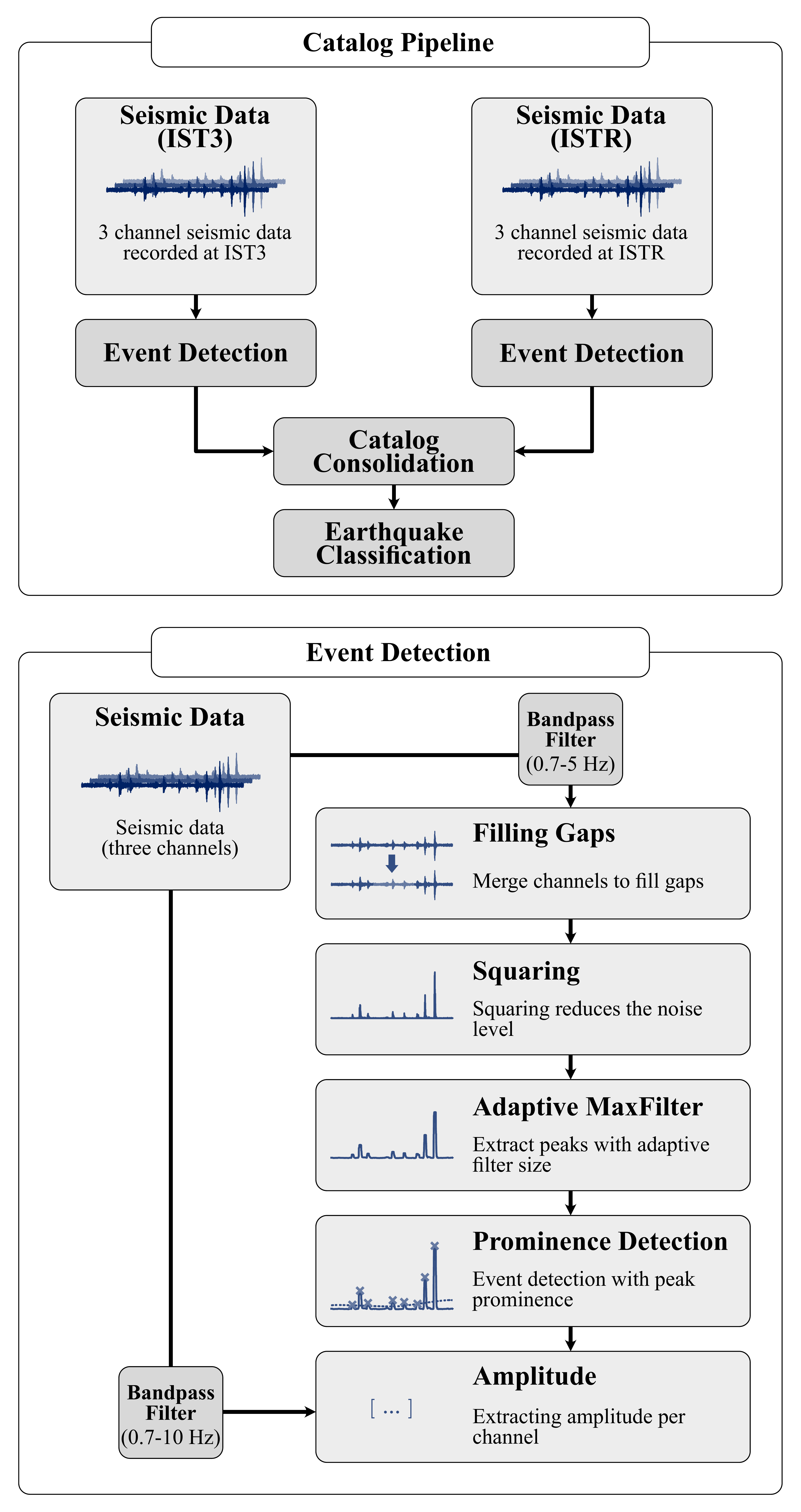}
    \caption{Overview of the catalog creation process using AWESAM including event detection, catalog consolidation and earthquake classification.}
    \label{fig:pipeline}
\end{figure}

Our proposed event detection approach consists of three steps which are outlined in the next sections. The algorithm compiles a catalog, containing the time and amplitude vectors for the three channels for all events. See Fig.~\ref{fig:pipeline} for an overview. AWESAM is implemented with the help of the SciPy\cit{scipy} ecosystem and obspy\cit{obspy}.

\subsection{Data sources - IST3 and ISTR on Stromboli}

AWESAM is developed for seismic data recorded at Stromboli, one of the most active volcanoes around the world. The volcano forms a 3 $\times$ 4~km island, located north of Sicily as part of the Aeolian Islands. In addition to the small and moderate eruptions (on average every 5 minutes), major explosions and paroxysms shake the island every few years in irregular time intervals: the last three paroxysms occurred in 2007 and then, after a long hiatus, in July and August of 2019\cit{precurs, majorexpl, jul19}.

The two seismic stations used by AWESAM are located on opposite sides of the island and have approximately equal distance to the crater area (\ref{fig:ExData}a). They are operated by the Istituto Nazionale di Geofisica e Vulcanologia (INGV) and the data is accessible through the INGV Seismological Data Center \cit{ingv}. The sampling rates are 100 Hz with three channels per station (HHN, HHE, HHZ).

Fig.~\ref{fig:ExData}b shows signals recorded at IST3 at randomly chosen dates. Both the noise level and the frequency of the eruptions vary greatly. The noise level, which is partly induced by ocean waves and wind\cit{windnoise}, obscures some small-amplitude events, making them difficult or impossible to detect. Hence it is a challenge for the algorithm to recognize as many of these small events as possible. 

\subsection{Event detection} \label{subsection:EvDet}

\subsubsection{Complementing channels} 

Channel HHN has the maximum amplitude for volcanic events in almost all cases considered. The reason is that the wavefield at the two stations is mainly horizontally polarized\cit{precurs}. Therefore it would be ideal to always use the HHN-channel. However, there are relatively frequent gaps in the recorded signals. In 2019 and 2020 (averaged over the three channels) 1~\% (IST3) and 11~ \% (ISTR), respectively, of the data are missing. It also turned out that gaps present in one channel are not always present in the other channels. Therefore, for the periods where HHN is missing, first, the HHE channel (having the second-largest amplitude) is used to fill gaps, and then, HHZ, which usually has the lowest amplitude. Hence, a stream with as few gaps as possible is generated. In this step, the channels are just composed, without scaling them with a factor, in order to preserve information about the polarization.

\begin{figure*}[tbp]
    \includegraphics[width=\textwidth]{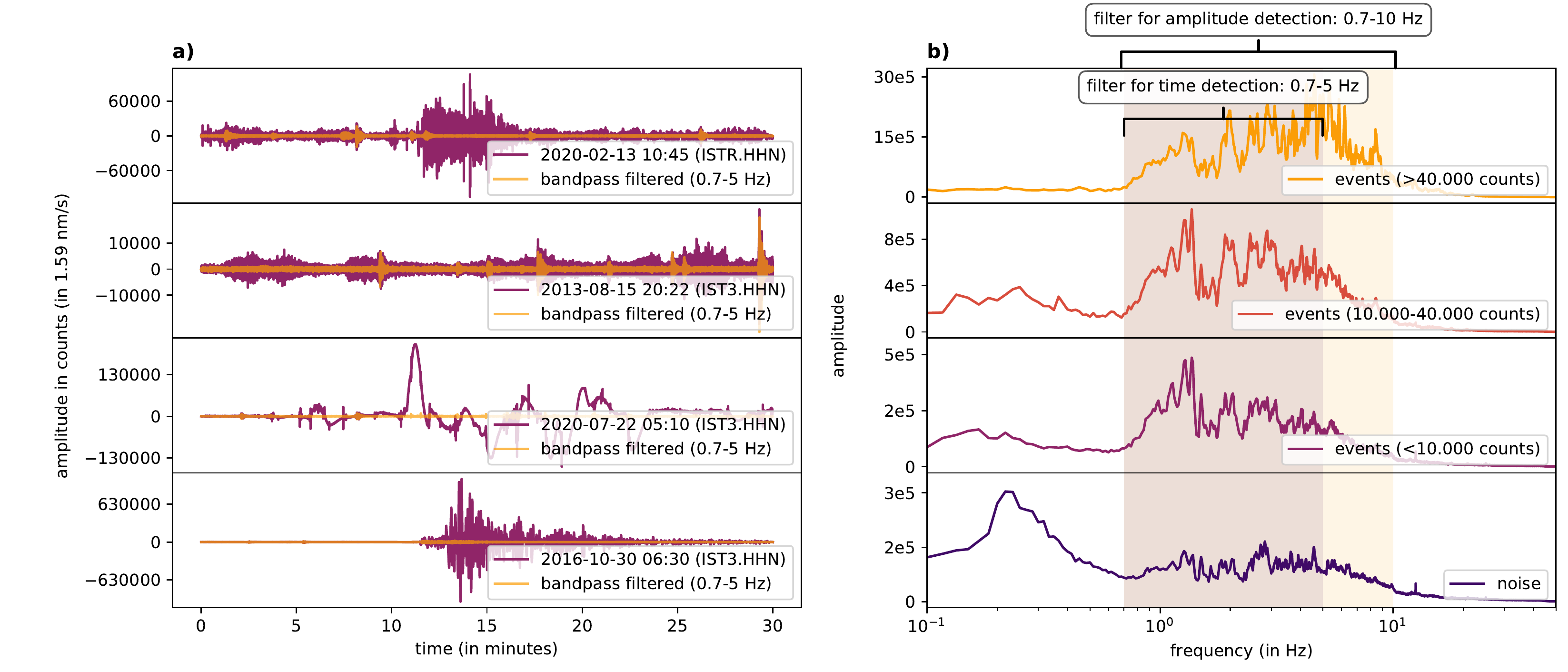}
    \caption{(a) Local disturbances at Stromboli. Their source is partly unknown. Because they can completely obscure the volcanic events, a band-pass filter (0.7-5 Hz) is applied, which is effective in removing the disturbances in most cases. 
    (b) Spectra of events with different signal amplitudes and noise levels (averaged spectra for 100 signals each). The intensity of noise is largest for frequencies near 0.2 Hz, probably due to ocean microseisms\cit{oceannoise}. The main frequency range for explosive events is between 0.7 and 10 Hz.}
    \label{fig:LocInter}
\end{figure*}

\subsubsection{Filtering}
After the continuous signal has been obtained, local disturbances in the data have to be excluded as best as possible. The source of the disturbances is partly unknown. Some may be caused by electrical malfunctions or maintenance/repair work. Fig.~\ref{fig:LocInter}a shows some examples, including an earthquake signal (6.5M, 500km from Stromboli). In the original signal, the disturbances may partially or completely cover the volcanic signals, making it impossible to detect them. However, the frequency spectrum of disturbances, which often ranges from 5-40 Hz, differs significantly from that of volcanic events (mostly 0.7-5 Hz, see Fig.~\ref{fig:LocInter}b).

Therefore we use a bandpass filter between 0.7 and 5 Hz. Fig.~\ref{fig:LocInter}a visualizes the recovered volcanic signal. Moreover, the earthquake signal has been filtered out as well. Note that this only works for distant, high magnitude earthquakes, because their frequencies are sufficiently low. The remaining earthquakes are identified using the method described in section~\ref{subsection:EqClass}.

However, this filter leads to a  10-40~\% change in the amplitude of eruptions. Unfortunately, there is no clear correlation between the filtered and unfiltered amplitude. For this reason, a slightly modified passband (0.7-10 Hz) is used to extract the amplitudes. This allows the resulting maximum amplitude to match the amplitude of the original signal well.

Incidentally, filtering is often used for general denoising. However, the volcanic events at Stromboli exhibit a similar frequency range as the ambient noise, making this method impractical. Therefore, the filtering is mainly used to exclude the (local) disturbances. Some papers also propose more complex filtering methods: for instance, all values in the time-frequency spectrogram, below a certain threshold,  are set to zero.\cit{denoise_fft} Still, these algorithms also require a sufficient difference in the frequency content of volcanic eruptions and noise. Finally, Machine Learning models can be used for denoising.\cit{denoise_unet, denoise_nn}. In this paper, we do not include a dedicated denoising step.

\subsubsection{Squaring} 
As already indicates, one challenge to detect eruptions is the varying noise level. Here, to reduce the noise level, the value of each sample in the time series data is squared. This step removes negative values in the data, but more importantly, it increases the prominence of peaks drastically. In theory, even higher exponents than 2 could be used. However, this would also make it increasingly difficult to identify smaller eruption events.

\subsubsection{Adaptive MaxFilter}
Generally, detecting peaks in the data can be achieved by applying a prominence threshold\cit{prominence}. At first, the oscillations in the signal must be removed in such a way that each event appears as a single peak. One possible solution is to compute the envelope of the data. However, events would still consist of multiple peaks. A better method is the application of a MaxFilter to the data. In this algorithm, a moving time window outputs the maximum value in each time step. By fine-tuning the window size (or filter size), this approach works reliably for most volcanic events.

\begin{figure*}[tbp]
    \includegraphics[width=\textwidth]{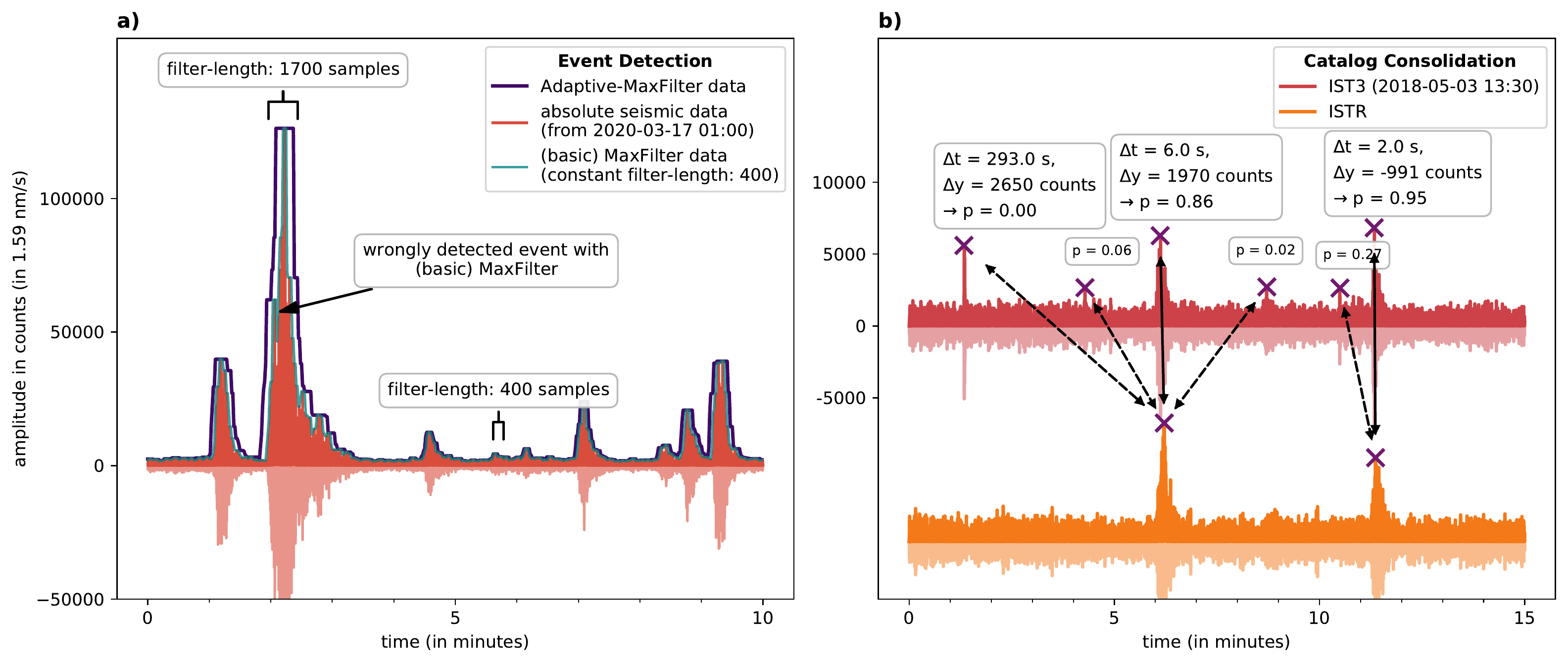}
    \caption{(a) Adaptive MaxFilter. The kernel size is adaptively computed from the mean of the absolute data, which enables to reliably detect the events of all amplitudes.
    (b) Catalog consolidation executed on example data to validate the generated catalogs with the signal from a second station. The arrows indicate the closest events detected at both stations, from which the distance and probability are computed. The dashed arrows denote that the two events do not correspond to each other. Note that the algorithm only uses the catalogs, the seismic data is just displayed for visualization.}
    \label{fig:explanation}
\end{figure*}

While this algorithm works well for events of smaller amplitude, it remains difficult to detect larger eruptions that have a substantially different shape and duration (i.e. usually the amplitude decay is much slower). To address this problem, the Adaptive MaxFilter algorithm is proposed which uses a variable filter size. For large explosions, the filter size automatically increases. Otherwise, the same event would be detected multiple times. Fig.~\ref{fig:explanation}a illustrates the differences between the two methods.

The filter size is determined from the average of the data in a surrounding time window (in this case with a length of 10 minutes). This approach results in kernel sizes between 300 and 2000 samples. If there is a paroxysm in the corresponding window, it can even increase to up to 10,000 samples. 

The stride, which determines the number of samples the time window moves in each step, should be set to 1 to get the best result. However, increasing the stride does not change the outcome of the algorithm substantially. Therefore, a stride of 100 samples is used (resulting in a sample rate of 1 Hz), causing a significant speed-up of the algorithm. Note that only the filter size changes in the Adaptive MaxFilter algorithm while the stride is kept constant.

\subsubsection{Prominence detection}
In the last step, the time and amplitude of the events are computed with a peak detection algorithm. Here, data that was filtered with a 0.7-10 Hz bandpass is used. We found that using the prominence of peaks as the threshold criterion is the best method to discriminate between eruptions and smaller peaks probably originating from background noise. (The prominence of a peak is the difference between its height and the lowest point to the next higher peak.\cit{prominence})

The changing prominence threshold is necessary to compensate for changes in the noise level. It is computed for every 10 minute time window (with the data from the same 10 minute time window):

\begin{equation}
    T = \alpha \cdot \frac {\mu_\text{data}}{\sigma_\text{data}}  \cdot \mu_\text{MaxFilter}
\end{equation}

\noindent where $\alpha = 1.5$ is a hyper-parameter that defines how many small peaks are detected and $ \mu_\text{MaxFilter}$ is the average of the data after applying the Adaptive MaxFilter algorithm in the corresponding time window. The ratio of the absolute original data ($\mu_\text{data}$) to the standard deviation ($\sigma_\text{data}$) can be interpreted as the noise-to-signal-ratio.

The result of the prominence detection is an array with the timestamps of all explosion events. The amplitude of each can be easily reconstructed from the adaptive MaxFilter-data (after computing the square root, to undo the squaring).

This algorithm allows the creation of solid event catalogs. However, some station-specific local disturbances that were not filtered out and tectonic earthquakes are still included. To be able to identify and remove these signals we developed the catalog consolidation algorithm.

\begin{figure}[tbp]
    \centering
    \includegraphics[width=0.9\columnwidth]{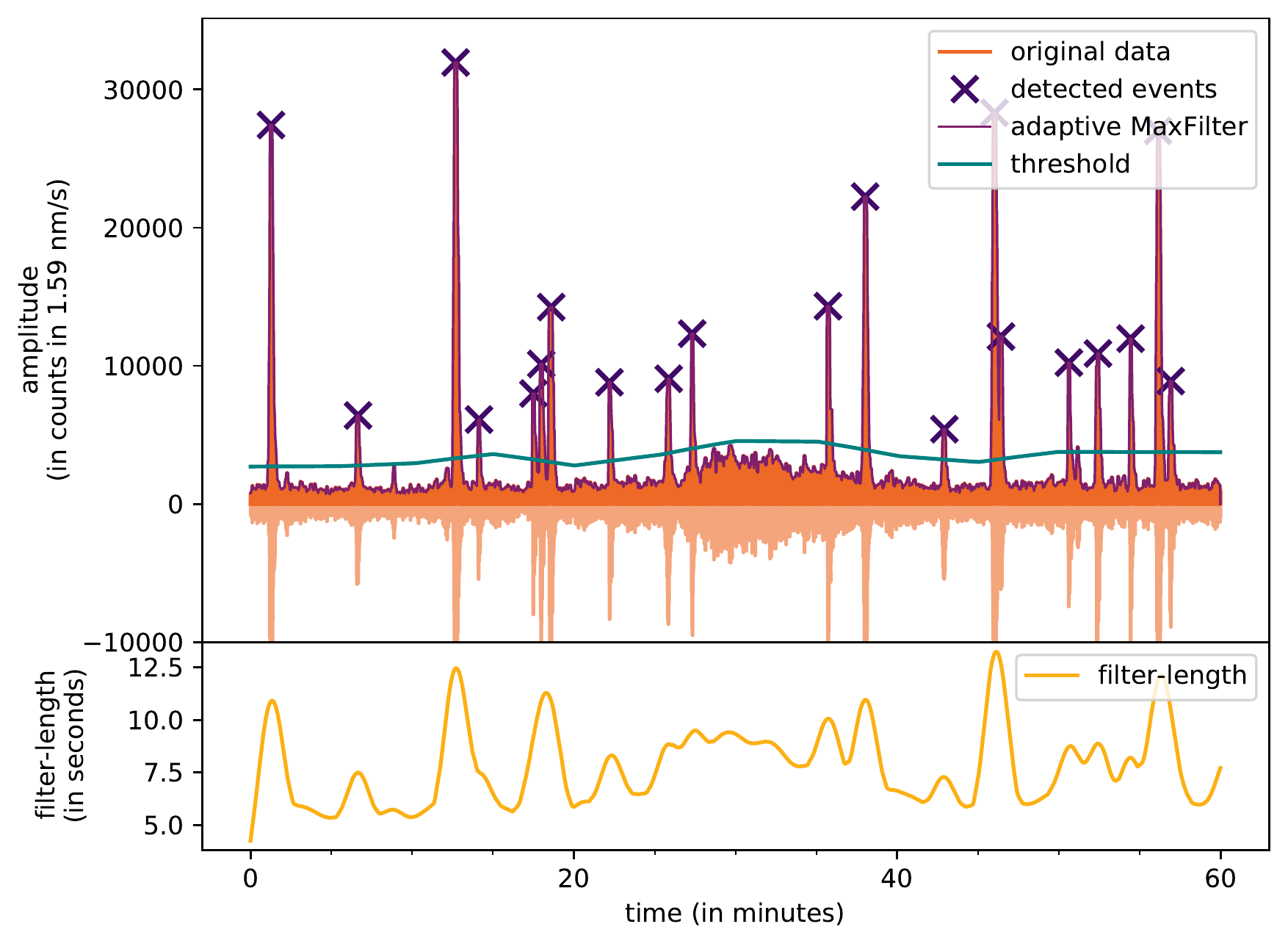}
    \caption{Event detection algorithm on example data. The adaptive prominence threshold and filter-size is shown.}
    \label{fig:EvDetEx}
\end{figure}

\subsection{Catalog consolidation}

The goal of this algorithm is to identify local disturbances (which occur at one single station) in a volcanic event catalog. The idea is that all significant events that occur at both stations must be non-local, thus being volcanic explosions or tectonic earthquakes.

As a basis, two catalogs (containing the time and amplitude of volcanic explosions) from two different stations are used. These two catalogs are computed independently with AWESAM's event detection algorithm (section \ref{subsection:EvDet}). In our case, one catalog is generated from the station IST3 (in the following called \textit{principal catalog}) and the other is obtained using data from ISTR (in the following called \textit{complementary catalog}). Note that the catalogs can be interchanged, which would lead to minor differences in the final catalog. In this algorithm, the catalog from IST3 is chosen to be the principle catalog since it has generally fewer local interferences and gaps, as found by manual inspection.

In the main step, for each event in the principal catalog, the ``closest" entry in the complementary catalog is selected. The closer this corresponding event is, the higher is the probability that its source is the volcano. If the distance is large,  then there is no corresponding event, which is an indication that the signal is not of volcanic origin. The distance between two events is defined as:

\begin{equation}
    d = \left \lVert g_\text{event}(y) \cdot\begin{pmatrix} \Delta t \\  \Delta y \end{pmatrix} \right \rVert
\end{equation}
\begin{equation}
    g_\text{event}(y) = \begin{pmatrix} 200/y & 0\\0 &0.1 /  y \end{pmatrix}
\end{equation}

\noindent where $\Delta t$ is the temporal difference and $\Delta y$ is the difference in amplitude. The metric $g$ is a 2x2-matrix that gives weights to the amplitude and time differences respectively. The metric was determined empirically, to represent a meaningful distance: for example, a 60 second difference in time should result in a higher distance $d$ than a 60 count difference in amplitude. 

Furthermore, the metric is a function of the amplitude $y$ of the current event from the principal catalog, because time and amplitude differences of corresponding events usually scale with the amplitude. For example, a 20 second time difference for a large volcanic event that lasts several minutes should not have a crucial influence, while for a small event it would not be negligible.

Thereby, the higher the distance, the lower the probability that an event's source is the volcano, and vice versa. This probability is calculated with 

\begin{equation}
    p = \exp ({-d})
\end{equation}

\noindent to normalize the distance to the interval $[0, 1]$.

Lastly, gaps in the principal catalog are complemented with events from the complementary catalog as best as possible: based on a gap table, which was generated during the detection of eruptions, all events from the complementary catalog that occurred during a gap in the principal catalog are copied to the principal catalog. Note that a gap is only considered if all three channels are missing at the same time.

This complements most gaps in the data, because usually at least one of the two stations delivers a signal in at least one channel. For example, in 2020 there exists no period where both stations have an outtake in all channels simultaneously.

Thereby, in all cases where only one station delivers a signal, no probability is assigned to the events. This is important to be able to distinguish events that could be consolidated and ones where there is no additional information available.

The catalog consolidation process, executed on example data, is shown in Fig.~\ref{fig:explanation}. If the amplitudes in both catalogs for corresponding events differ (due to different instruments, for example) a unit conversion is needed. For Stromboli, however, this is not necessary, because the same instruments\cit{station_ist3, station_istr} are used and the stations have equal distance to the crater. 

\begin{figure*}[tbp]
    \includegraphics[width=\textwidth]{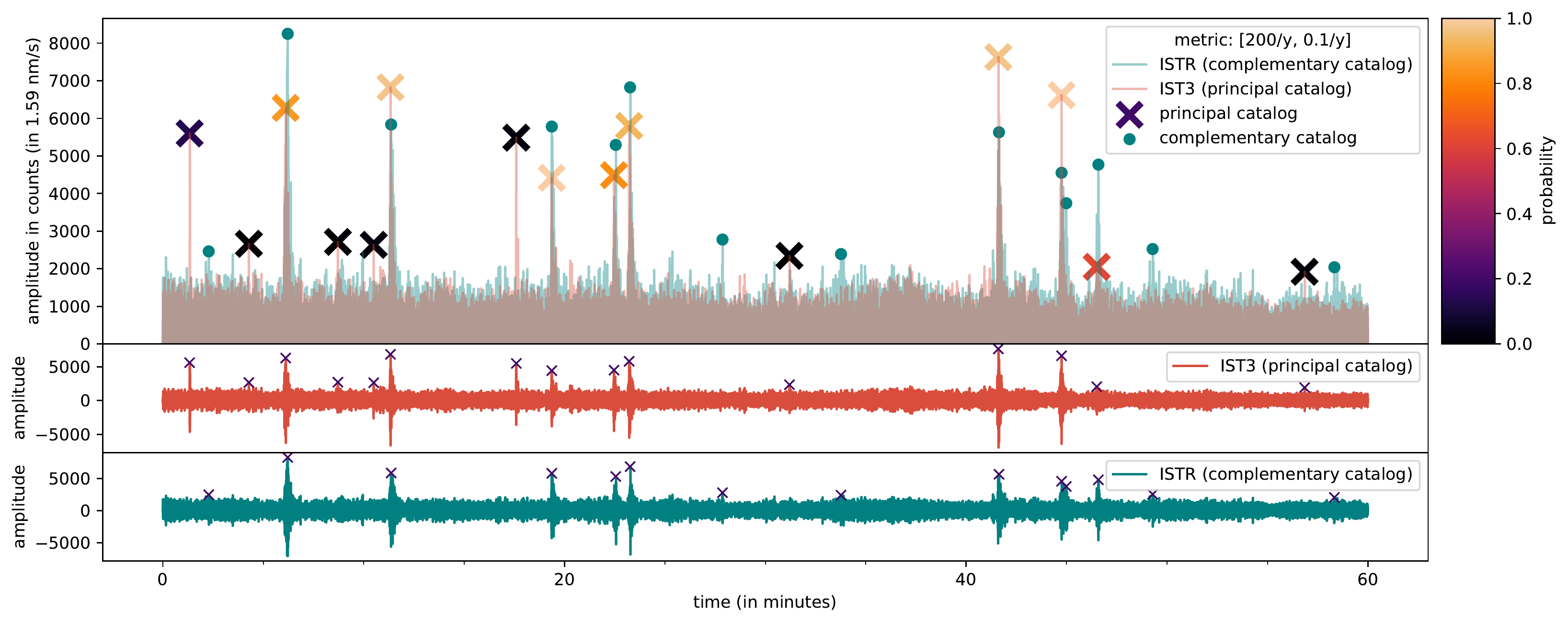}
    \caption{Catalog consolidation test. Events that only appear at one of the two stations are assigned a small probability (that the source of the signal is the volcano). Events that match in both catalogs have a probability close to 1.}
    \label{fig:CatCons}
\end{figure*}

This algorithm effectively removes station-specific disturbances from the catalogs. However, if the algorithm is not correctly configured it can also assign low probabilities to volcanic events. This happens when an event is only recognized in the principal catalog.

\subsection{Earthquake classification} \label{subsection:EqClass}

\begin{figure}[tb]
    \includegraphics[width=\columnwidth]{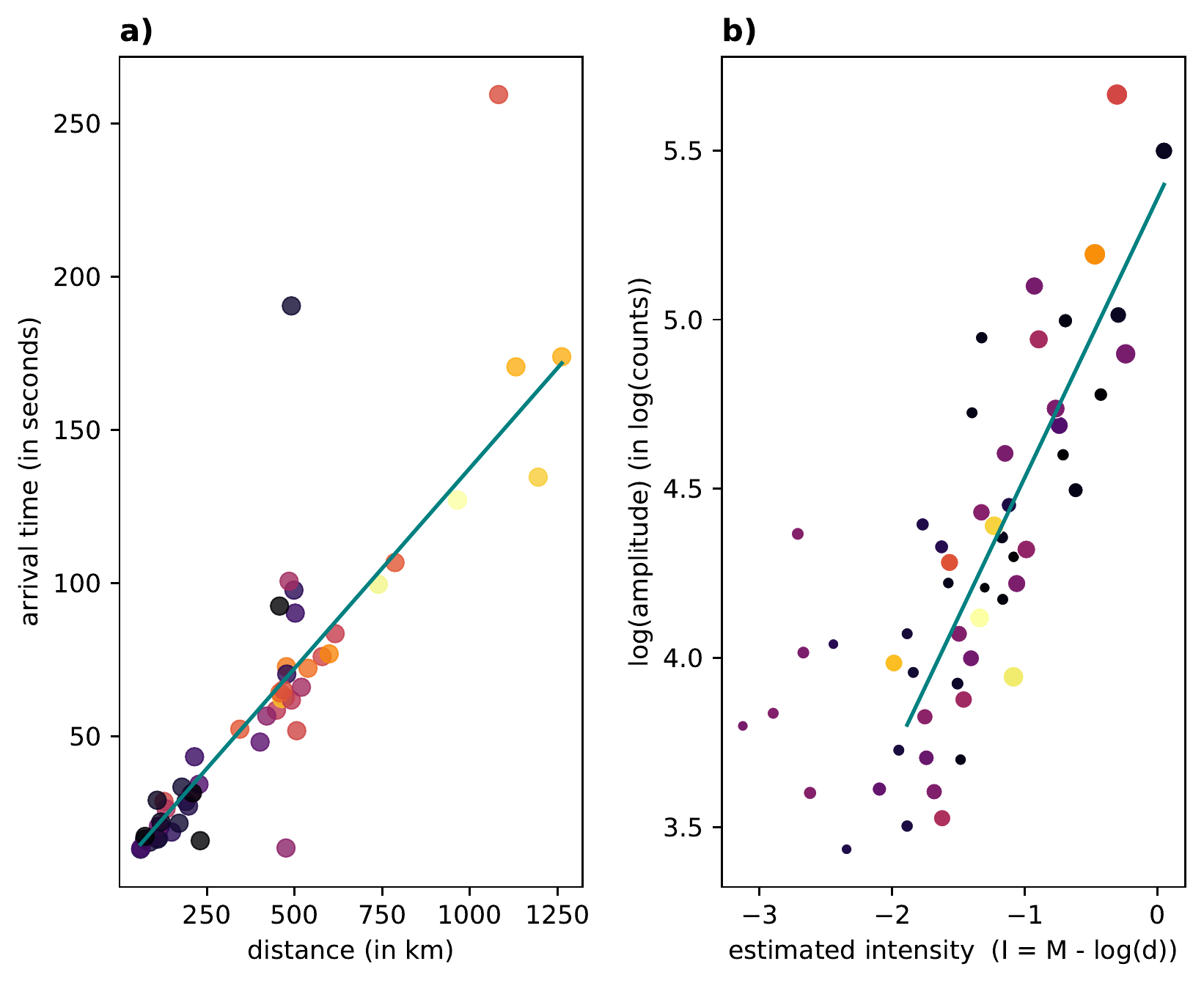}
    \caption{Estimation of arrival time and amplitude of tectonic earthquake signals recorded at Stromboli.}
    \label{fig:EqEstim}
\end{figure}

While local disturbances are accounted for during catalog consolidation, tectonic earthquakes are not identified. As shown in Fig.~\ref{fig:LocInter}a, bottom panel, only distant earthquakes may be filtered out. To address the remaining cases, the official earthquake catalog from INGV can be utilized. One problem is that this catalog does not include information about the arrival time and amplitude of an earthquake at Stromboli, but only information about its origin time and hypo-center. To identify events in the seismic data as earthquakes, the local arrival time and amplitude have to be estimated.

To estimate these two parameters, we determine the arrival time and amplitude manually for several dozen larger earthquakes, which can be clearly seen in the seismic data. Based on this, as expected, a correlation between distance and arrival time can be seen clearly in Fig.~\ref{fig:EqEstim}a. The proportionality constant is $6.6~\frac{\mathrm{km}}{\mathrm{s}}$, which corresponds to the average crustal P-wave velocity in the area. The arrival time is therefore given by:

\begin{equation}
    t = T + \frac d {6.6~\text{km}/ \text{s}}
\end{equation}

\noindent where $T$ is the time of the earthquake at the hypo-center and $d$ is the distance of the hypo-center to the craters of Stromboli. Note that the distance to Stromboli takes the focal depth of earthquakes into account, but not the curvature of the earth.

In addition, an intensity measure is defined, which is based on a method introduced by Pasolini et al.\cit{eqatten}:

\begin{equation}
    I = M - \ln(d)
\end{equation}

Here the natural logarithm is used. The correlation between the intensity and the actual amplitude (Fig.~\ref{fig:EqEstim}b) is sufficient to roughly estimate if an earthquake will be visible at Stromboli or not. The relationship between the intensity  $I$ and the amplitude in counts $a$ is:

\begin{equation}
    a = \text{10} ^ {\text{0.83} \cdot I + \text{5.36}} 
\end{equation}

With this information, the catalog consolidation algorithm can be used to compute the probability that an event is an earthquake. One difference to the previous application (the purpose of which was to identify local disturbances) is that the earthquake catalog is now the complementary catalog. Also, a different metric which was again found empirically is used, because, firstly, the estimation of the arrival time is not very accurate and, secondly, because earthquakes usually have a duration of several minutes (compared to 40 seconds for the average volcanic event):

\begin{align}
    g_\text{\text{earthquake}}(y) &= 0.1 \cdot g_{\text{event}}(y) \nonumber\\
    &= \begin{pmatrix} 20/y & 0\\0 &0.01 /  y \end{pmatrix}
\end{align}

After this, the event catalog is complete and includes two probabilities (volcanic event-probability \& earthquake probability) which provide more information about the source of the events. Please note that the resulting probabilities are not meant to exclude events completely. Instead, they should facilitate the later analysis, by providing additional information.

\section{Performance evaluation and testing}

\begin{figure}[tbp]
    \includegraphics[width=\columnwidth]{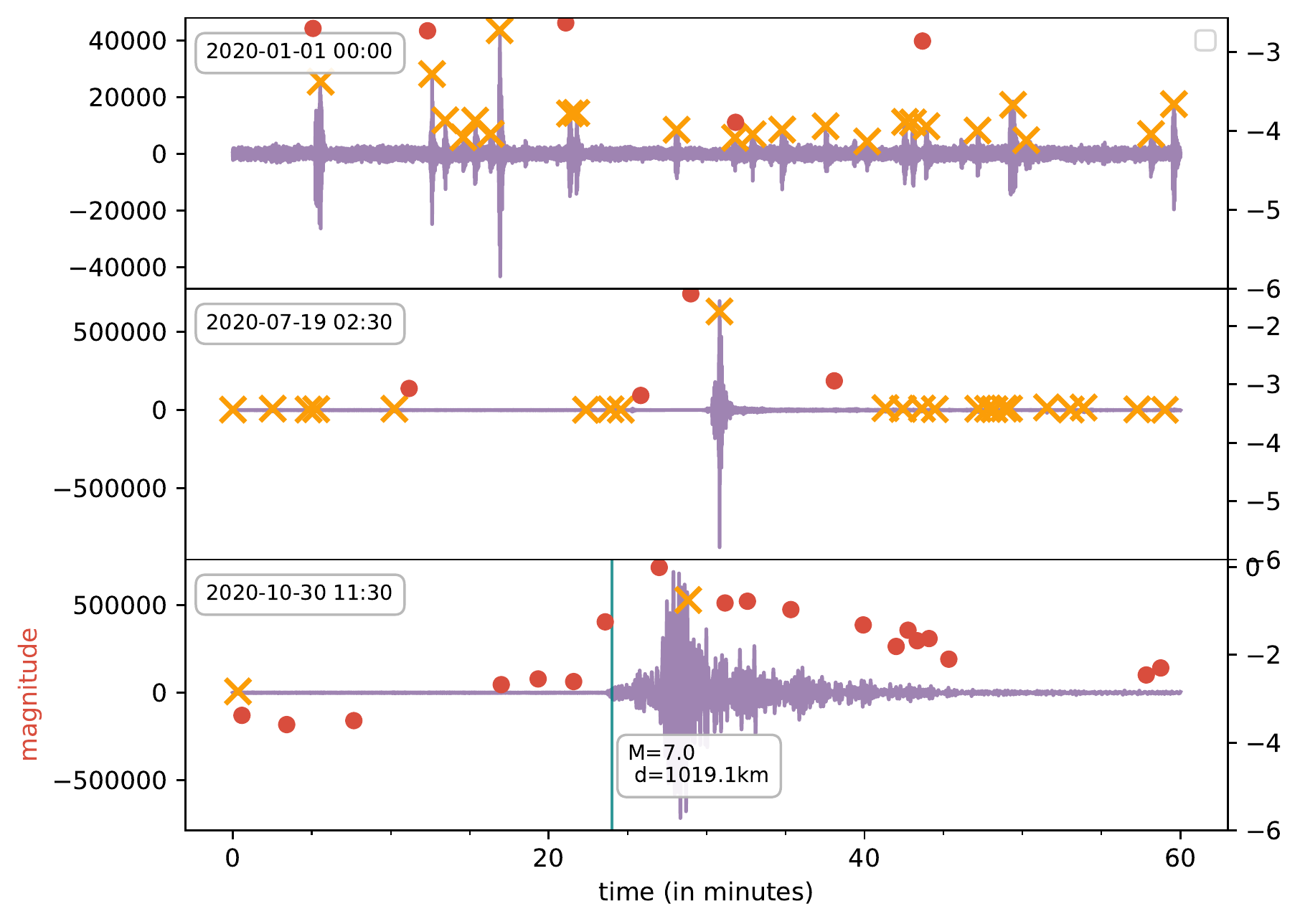}
    \caption{Comparison AWESAM-catalog with EOLO-catalog\cit{eolo}. The catalogs are very different. Larger events appear in both, as shown in the second plot. Earthquakes (lower plot), that are excluded by AWESAM, are present in Eolo}
    \label{fig:Eolo}
\end{figure}

In order to evaluate the quality of the AWESAM-catalog, a reference database is needed. INGV provides a catalog of very-long-period (VLP) or explosive events at Stromboli.\cit{eolo} It is based on  EOLO\cit{eolopaper}, an automated analysis of recordings from the seismic network on Stromboli, and contains information about the time, magnitude, location and depth of VLP events.

However, the entries of the EOLO catalog show little agreement with the results from our analysis. Fig.~\ref{fig:Eolo} depicts some examples. The second plot includes a larger event, which cannot be traced back to a tectonic earthquake, which is not picked very accurately by Eolo. In addition, the third plot in this figure features a signal which is likely caused by an earthquake. While these events would be excluded in the AWESAM catalog, they are included by EOLO.

To compare our catalog, we created a catalog of events by hand, containing over 4500 events, selected from 7.5 days of randomly chosen data. For this purpose, we developed the EventPicker software, a sub package of AWESAM. It simplifies event selection down to a single click with the mouse. 

To compute the accuracy, these two catalogs (AWESAM-catalog and hand-picked catalog) can be compared using a modification of the catalog consolidation algorithm: In the first step, the probabilities of all AWESAM events (in the time range of the hand-picked events), which now form the principal catalog, are calculated, using the hand-picked catalog as the complementary catalog. Next, by applying a catalog consolidation vice versa, using the hand-picked catalog as the principle and the AWESAM-catalog as the complementary catalog, we generated a measure of how many events in one catalog are contained in the other. In consequence, there are two accuracy measures, from which an overall accuracy can be derived:

\begin{itemize}
    \item $\pmb{A_1}$: Measures the fraction of events in the AWESAM-catalog that are also contained in the hand-picked catalog. This value is the average of the probabilities (computed with the catalog consolidation algorithm), while using AWESAM as the principal catalog and the hand-picked as the complementary catalog.
    \item $\pmb{A_2}$: Measures the fraction of events in the hand-picked catalog that are also contained in the AWESAM-catalog. This value is the average of the probabilities (computed with the catalog consolidation algorithm), while using AWESAM as the complementary catalog and the hand-picked as the principal catalog.
    \item \textbf{Overall accuracy:} $A = \frac 12 (A_1 + A_2)$
\end{itemize}

\begin{figure}[tbp]
    \includegraphics[width=\columnwidth]{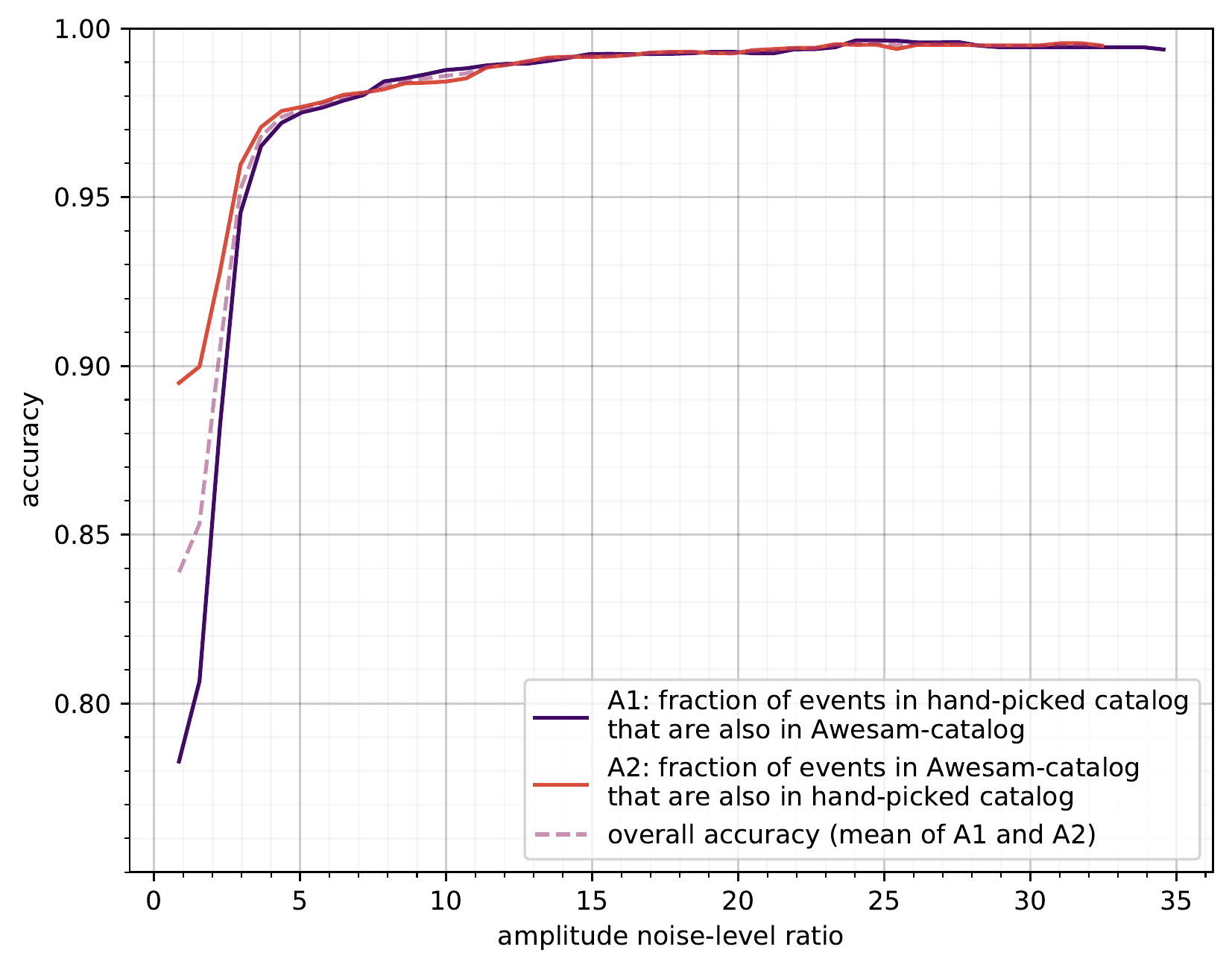}
    \caption{The accuracy curves show how the accuracy depends on the signal-to-noise ratio. The more prominent the peak over the noise level, the higher the probability that it is detected.}
    \label{fig:Acc}
\end{figure}

Note that the specific values of the accuracies depend on the metric used in the catalog consolidation. Fig.~\ref{fig:Acc} illustrates the accuracy curves of the catalogs. These curves reveal the dependence between the accuracy and the amplitude of events relative to the noise level (signal-to-noise ratio -- SNR): The higher the SNR, the higher the probability that events are detected. For example, events with an SNR larger than three are detected with an accuracy of 95~\%. Thereby $A_1$ and $A_2$ hold different information: The fact that $A_1$ is for almost all signal-to-noise ratios smaller than $A_2$ shows that the hand-picked catalog contains some events that do not appear in the AWESAM catalog, but not the other way around (almost all events in the AWESAM catalog are also in the hand-picked catalog). However, Fig.~\ref{fig:Acc} also exhibits that this only happens for events, whose amplitudes are close to the noise level. The noise level is computed with the 95th percentile of the absolute seismic data.

\section{Catalog analysis - amplitude-frequency relation}

The catalog offers a wide range of possibilities for further analysis. For example, amplitude-frequency-relations can be investigated and a detailed inter-event-time analysis could help understanding possible regularities in occurrences of explosive events. The latter was observed by Bevilacqua et al.\cit{majorexpl} for major explosions and paroxysms. Natural time analysis introduced by Varotsos et al.\cit{crittime, nattime} is also a promising approach to investigate the volcanic activity. Additionally, it can serve as an indicator for the volcanic activity.

Particularly, the catalog is useful for any analysis that relies on the data of explosive events over long periods of time i.e. years and decades. In earlier investigations, for example by Nishimura et al.\cit{gutrichter}, the amplitude-frequency relationship was derived based on data for two months only. 

\begin{figure}[tb]
    \includegraphics[width=\columnwidth]{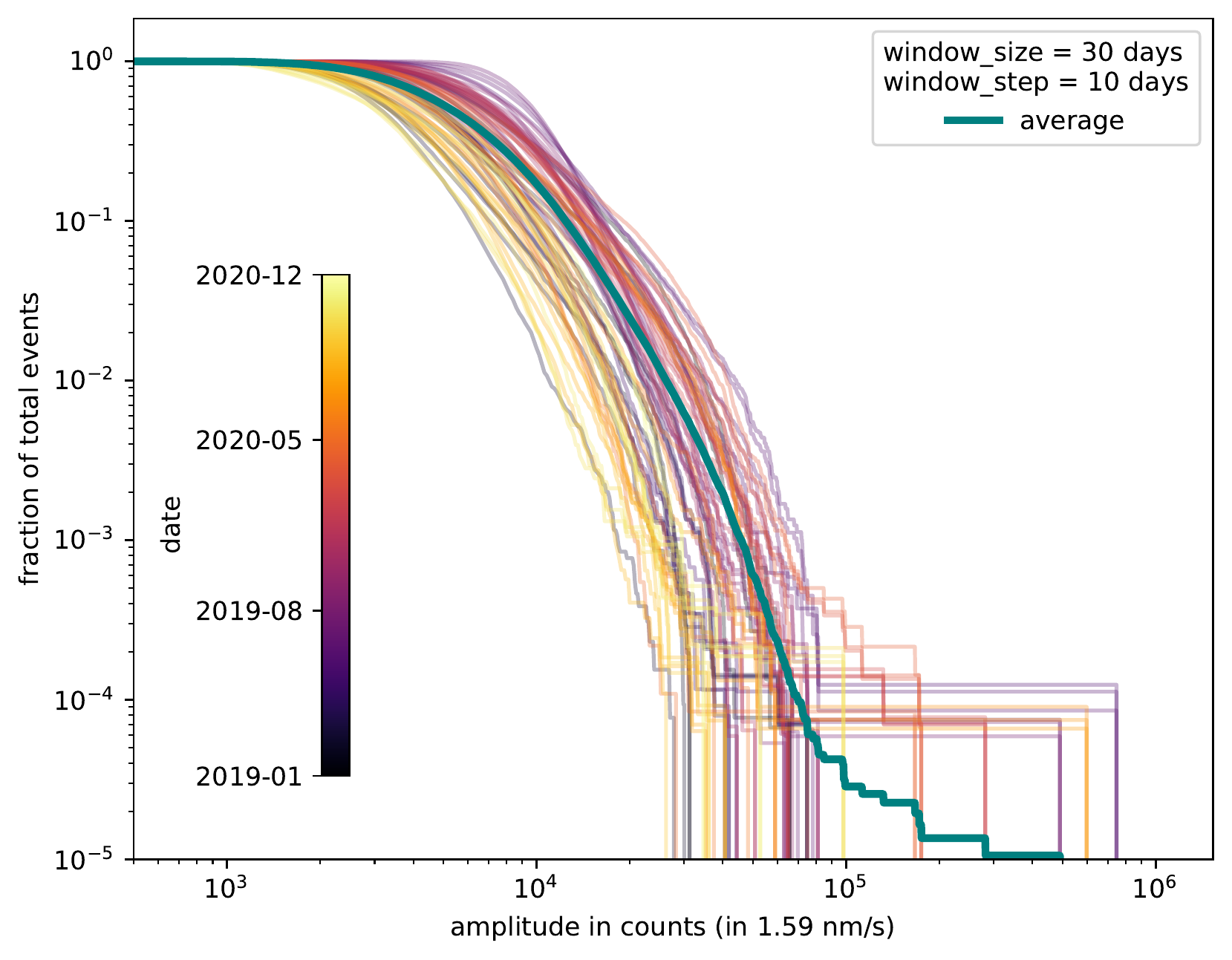}
    \caption{The amplitude-frequency relation and its temporal development over a period of two years (2019-2020) at Stromboli. The catalog contains more than 290.000 events for this time period. The window size used to compute the individual curves is 30 days; a window step size of 5 days was used.}
    \label{fig:MagFreq}
\end{figure}

As a first example for the application of the information provided by the catalog, the temporal evolution of the amplitude-frequency relationship for the year 2020 is shown in Fig.~\ref{fig:MagFreq}. In contrast to results reported earlier\cit{gutrichter}, the curve based on all events (labeled ``average") is characterized by a significant change in slope for events with amplitudes below and above 50.000 counts. This may indicate that the larger events are due to different physical mechanisms than the smaller, more frequent explosions. However, the change in slope is only supported by only $\sim$10 events. To draw further conclusions, the catalog needs to be extended further to cover years and thus events. However, more detailed analyses of the catalog and these observations will be presented in a forthcoming paper.

\section{Conclusions}

We introduced a detection algorithm for detecting small and large seismic events due to volcanic explosions at Stromboli. To further improve the event catalog we developed the catalog consolidation algorithm that removes local disturbances and tectonic earthquakes by including data from a second station and by using information derived from the official earthquake catalog of INGV.
The resulting event catalog of volcanic explosions is almost complete and cleaned up and will be used for further studies of the eruptive behavior at Stromboli. This module can further be adapted to generate catalogs for other volcanoes and also to generate earthquake catalogs by using raw seismic recordings.

\section*{Acknowledgment}
This research is supported by the ``KI-Nach\allowbreak wuchswissenschaftlerinnen" - grant SAI 01IS20059 by the Bundesministerium für Bildung und Forschung - BMBF. Calculations were performed at the Frankfurt Institute for Advanced Studies' new GPU cluster, funded by BMBF for the project Seismologie und Artifizielle Intelligenz (SAI).

\section*{Competing Interests}
The authors declare no competing interests.

\bibliographystyle{bst/elsarticle-num}
\bibliography{references.bib}

\end{document}

%% file: authors.tex
\author[fias,unimainz]{Darius Fenner}
\author[fias,geo]{Georg Rümpker}
\author[fias]{Wei Li} 
\author[fias,geo]{Megha Chakraborty} 
\author[fias,phys]{Johannes Faber} 
\author[fias,geo]{Jonas Köhler} 
\author[fias,phys,gsi]{Horst Stöcker} 
\author[fias]{Nishtha Srivastava} 
\ead{srivastava@fias.uni-frankfurt.de}

\affiliation[fias]{
    organization = {Frankfurt Institute for Advanced Studies},
    postcode = {60438},
    city = {Frankfurt am Main},
    country = {Germany}
}

\affiliation[unimainz]{
    organization={Johannes Gutenberg-Universität Mainz},
    postcode = {55122},
    city = {Mainz},
    country = {Germany}
}

\affiliation[geo]{
    organization = {Institute of Geosciences, Goethe-University Frankfurt},
    postcode = {60438},
    city = {Frankfurt am Main},
    country = {Germany}
}

\affiliation[phys]{
    organization = {Institute for Theoretical Physics, Goethe Universität},
    postcode = {60438},
    city = {Frankfurt am Main},
    country = {Germany}
}

\affiliation[gsi]{
    organization = {GSI Helmholtzzentrum für Schwerionenforschung GmbH},
    postcode = {64291}, 
    city = {Darmstadt},
    country = {Germany}
}